\documentclass[runningheads]{llncs}
\pdfoutput=1

\usepackage{graphicx}
\usepackage{subfig} 
\usepackage{float}
\usepackage{color}
\usepackage{tikz}
\usetikzlibrary{shapes,arrows,patterns,positioning} 
\usepackage{caption}
\usepackage[normalem]{ulem}

\usepackage{comment}
\usepackage{multirow}
\usepackage[ruled,linesnumbered]{algorithm2e}
\usepackage{amssymb}
\usepackage{amsmath}
\usepackage{empheq}
\usepackage{ulem}

\usepackage{breakcites} 

\usepackage{xspace,tcolorbox}
\usepackage{hyperref} 
\usepackage[utf8]{inputenc}
\usepackage{textcomp}

\usepackage[symbol]{footmisc}

\usepackage[flushleft]{threeparttable}

\tikzstyle{decision} = [diamond, draw, 
    text width=4.5em, text badly centered, node distance=3cm, inner sep=0pt]
\tikzstyle{block} = [rectangle, draw, 
    text width=5em, text centered, rounded corners, minimum height=4em]
\tikzstyle{line} = [draw, -latex']
\tikzstyle{cloud} = [draw, ellipse,fill=red!20, node distance=3cm,
    minimum height=2em]
    \newenvironment{cppalgorithm}[1][htb]
  {\renewcommand{\algorithmcfname}{Program}
   \begin{algorithm}[#1]%
  }{\end{algorithm}}

\newcommand{\revise}[1]{{\color{black}#1}}
\newcommand{\REM}[1]{}
\REM{ multiline
comment }

\usepackage[formats]{listings}
\usepackage{xcolor}

\usepackage[left=3.2cm, top=3.8cm, asymmetric]{geometry} 

\begin{document}




\title{SCOPE: Secure Compiling of PLCs in Cyber-Physical Systems}

\author{Eyasu Getahun Chekole{\inst{1}\textsuperscript{,}\thanks{\textit{This work was partly done when E.G. Chekole was a Ph.D. student at the Singapore University of Technology and Design.}}} \and Mart\'in Ochoa\inst{2,3} \and Sudipta Chattopadhyay \inst{3}}
%

\institute{Institute for Infocomm Research, A*STAR, Singapore 138632, Singapore\\
\and Singapore University of Technology and Design, Singapore 487372, Singapore\\
\and AppGate Inc, Bogot\'a, Colombia
}
\maketitle


\begin{abstract}
Cyber-Physical Systems (CPS) 
are being widely adopted in critical infrastructures, such as smart grids, 
nuclear plants, 
water systems, transportation systems, manufacturing and 
healthcare services, among others. However, the increasing prevalence of cyberattacks targeting them raises a growing security concern in the domain. In particular, memory-safety attacks, that exploit memory-safety vulnerabilities, constitute a major attack vector against real-time control devices in CPS. 
Traditional IT countermeasures against such attacks have limitations when applied to the CPS context: they typically incur in high runtime overheads; which conflicts with real-time constraints in CPS and they 
often abort the 
program when an attack is detected, thus harming availability of the system, which in turn can potentially result in damage to the physical world.
In this work, we propose to enforce a \emph{full-stack} memory-safety (covering \emph{user-space} and \emph{kernel-space} attack surfaces) based on secure compiling of PLCs to detect 
memory-safety attacks in CPS. 
Furthermore, to ensure availability, we enforce a resilient 
mitigation technique that bypasses illegal memory access instructions at runtime by dynamically instrumenting low-level code. 
We empirically measure the computational overhead caused by our approach on two experimental settings based on real CPS. 
The experimental results show that our approach effectively and efficiently detects and mitigates memory-safety attacks 
in realistic CPS. 

\end{abstract}

\keywords{
Critical Infrastructures Security \and 
CPS Security \and Software Security \and Memory Safety \and  Efficiency \and Resilience \and Availability 
}




\section{Introduction}\label{introduction} 


Cyber-physical systems \cite{shacps,lee_introductiontocps,lee_cpsdesignchallenges} 
are being widely adopted in various mission critical infrastructures including smart grids, water treatment and distribution systems, transportation, nuclear plants, robotics and  manufacturing, among others. Despite their importance in such critical infrastructures, the increasing cyberattacks 
targeting them 
poses a 
growing security concern. 
\revise{
One important class of cyberattacks in CPS are memory-safety attacks~\cite{sok,memory_safety_attacks_survey} that target programmable logic controllers (PLCs). 

A typical PLC consists of three main software components -- the PLC firmware, the control software (i.e. the control logic) and the underlying OS hosting the PLC. Since 
these software components are commonly implemented in C/C++ languages 
(for the sake of efficiency), 
they are susceptible to memory-safety vulnerabilities, such as buffer overflows, use-after-free errors (dangling pointers), use-after-return errors, initialization order bugs, and memory leaks. Consequently, a wide-range of these vulnerabilities are being regularly discovered even in modern PLCs
~\cite{cve_ab,cve_ab2,cve_ab3,cve_siemens1,cve_siemens2,cve_sem1,cve_sem2,cve_abb} 
and Linux kernels~\cite{cve_linux}.

These vulnerabilities could lead to runtime crashes, 
which 
can severely affect safety- and availability-critical systems, such as CPS. More importantly, these vulnerabilities can also be exploited by 
memory-safety attacks. Memory-safety attacks, such as code-injection~\cite{code_injection_attacks} and code-reuse~\cite{code_reuse_attacks1} attacks, can corrupt the memory system of a vulnerable program to hijack or subvert its operations. In CPS, these attacks can target the PLC's firmware and control software (\emph{user-space}) or the underlying OS kernel hosting the PLC (\emph{kernel-space}). Therefore, both the runtime crashes and memory-safety attacks are critical concerns in CPS.

}



\revise{
To overcome 
the runtime crashes and security challenges, a wide-range of countermeasures, often referred as \textit{memory-safety tools}, have been 
developed 
\cite{asan,kasan,softbound,cets,ccured,Mudflap,memsafe,ropocop,cup,safecode,safedispatch,cfi,cfi_cots,cfi_gcc,cfi_sp,dieharder1,dieharder2,deputytinyos,deputy,nesc,failure_oblivious,cfi_data-attacks}. 
However, the hard real-time and availability requirements imposed in CPS, alongside the use of resource-constrained edge devices, 
limit the practical applicability of certain 
memory-safety tools available. This is because, the high memory-safety overheads (MSO) induced by certain memory-safety tools compromise the real-time requirements in CPS. Furthermore, the non-resilient mitigation strategies exerted in certain memory-safety tools (e.g. plainly aborting/restarting the victim system when a memory-safety attack is detected) compromise availability of the system. Therefore, the \emph{efficiency} and \emph{mitigation resilience} of memory-safety tools are crucial requirements in CPS that should be met alongside strong security guarantees. 
}


\noindent \emph{Efficiency} --
\revise{Most memory-safety 
tools, especially the code-instrumentation ones incur in high runtime overheads, e.g., RopoCop~\cite{ropocop} (240\%), CUP~\cite{cup} (158\%), CCured~\cite{ccured} (150\%), SoftBoundCETS~\cite{softbound} (116\%), and MemSafe~\cite{memsafe} (87\%). 
This high overhead may unacceptably compromise performance of the CPS. If the CPS 
real-time} constraints are not met, major consequences can follow such as disruption of the control-loop stability, incorrect control by the use of stale information, availability issues, system damage (in the worst case), etc.
\revise{
Thus, the trade-off between security 
and 
efficiency remains as one of the main conflicting design challenges in CPS.} 

\noindent \emph{Mitigation resilience} -- 
\revise{Most of the existing memory-safety 
tools} do not have a resilient mitigation strategy. They are primarily designed to abort or reboot the victim system \revise{when a memory-safety attack or violation 
is detected, thus leading to system unavailability. 
Such ineffective mitigation strategies are not acceptable in systems with stringent availability 
requirements, such as CPS}. Because, system unavailability in CPS leaves the control system into an unsafe state and leads disruption of 
the CPS dynamics, which may result in a complete system 
failure (cf. Section \ref{modeling_physical_state_resiliency}). Thus, system availability is also a critical requirement 
in CPS. 

\revise{
Therefore, vis-a-vis 
the real-time and availability requirements (which we particularly associate with the efficiency and mitigation resilience of the security solutions) are equally critical as the runtime crashes and security concerns in the CPS environment. 
}

\paragraph{Our approach}
To address these challenges, we propose a countermeasure called 
{\bf S}ecure {\bf C}ompiling {\bf O}f {\bf P}LCs in cyb{\bf E}r-physical systems ({\bf SCOPE}). 
Inspired by our recent work, CIMA \cite{cima_cose}, our approach is based on the intuition of proactively stopping memory-safety violations from happening, thereby preventing both runtime crashes and 
memory-safety attacks in the process. To accomplish this, we follow a compile-time code-instrumentation based approach that offers stronger guarantees in terms of error coverage and detection accuracy, despite introducing higher performance overheads. To cover the attack surfaces in user-space and kernel-space, we escalate our solution to a full-stack memory-safety countermeasure, comprising a user- and kernel-space memory-safety 
solutions. 

After researching over several available tools, we port the popular memory-safety tools, such as AddressSanitizer(ASan)~\cite{asan} and Kernel Address Sanitizer (KASan)~\cite{kasan}, as a user-space and kernel-space memory error detector tools, respectively, by fixing their limitations to work in a CPS environment. 
We enhance this detection strategy 
by integrating our recent mitigation work, CIMA\cite{cima_cose}, that systematically combines a compile-time code-instrumentation and runtime monitoring techniques to 
resiliently mitigate the detected memory-safety attacks.  

This work is an extension of our previous works. \revise{In brief, it combines our prior memory-safety works on: 1) \emph{a user-space attack detection}~\cite{eyasu_cybericps} (disregarding availability attacks (i.e. no mitigation resilience) and without considering a kernel-space memory-safety); 2) \emph{a user-space mitigation resilience}~\cite{cima_cose} 
(without considering a kernel-space memory-safety); and 3) \emph{a kernel- and user-space attack detection}~\cite{eyasu_essos} (disregarding availability attacks).} So, 
in this work, we integrated the three approaches together to form a full-stack memory safety. 
Although we previously studied them 
separately, 
ultimately they should 
all together be part of the secure compilation strategy. \revise{It is an open question whether a practical CPS would tolerate the joint computational overhead induced by the proposed full-stack memory-safety. 
} 

\paragraph{Evaluation} The effectiveness of the proposed full-stack memory-safety is experimentally evaluated. Our experiments are based on two realistic CPS testbeds: SWaT (Secure Water Treatment System)~\cite{swat} and SecUTS (Secure Urban Transportation System)~\cite{secuts_paper}, comprising  real-world vendor-supplied PLCs.  However, the vendor's PLC firmware (both in SWaT and SecUTS) is closed-source, hence we could not incorporate our memory-safety solutions in these PLCs. To circumvent this challenge, we prototyped our experimental testbeds, which we call open-SWaT and Open-SecUTS, using open-source PLCs to mimic the behavior of SWaT and SecUTS, respectively, according to their detailed operational profiles. Then, we report experiments conducted on Open-SWaT and Open-SecUTS.

A strong memory-safety countermeasure apparently incurs high cost, i.e., performance overhead, 
which might not be acceptable in CPS due to the hard real-time and availability requirements imposed in these systems. 
To evaluate the acceptability of such overheads, we briefly modeled the CPS design constraints, such as the \emph{real-time} and \emph{physical-state resiliency} requirements. These models aid as benchmarks to evaluate tolerability of the performance overheads and resilience of the system dynamics in CPS. 
Subsequently, we evaluated the effectiveness of our full-stack memory-safety solution on the Open-SWaT and Open-SecUTS testbeds. In particular, we evaluate tolerability of the induced performance overhead in accordance with the CPS real-time constraints we modeled. Our experimental results on Open-SWaT and Open-SecUTS reveal that the introduced memory-safety overhead of 91.02\% (for Open-SWaT) and 85.49\% (for Open-SecUTS) would not impact the normal operations of SWaT and SecUTS. \revise{Furthermore, 
our user-space mitigation strategy also meets physical-state resiliency of the CPS testbeds under test. }

In general, our full-stack countermeasure efficiently and successfully prevents m\-e\-m\-o\-r\-y-safety violations from happening without compromising availability of the system. To the best of our knowledge, this is not achieved by any prior work. Although our proposed 
memory-safety is applicable for any computing system 
involving C/C++ programs, we 
particularly focused on the CPS domain in this research. This is because, unlike the mainstream systems, CPS often imposes conflicting design constraints including real-time guarantees and physical-state resiliency -- involving its physical dynamics and security. Note that attacks that manipulate sensor or actuator values at storage or communication levels are out of our scope, and can be handled via orthogonal approaches, e.g., using physics-based approaches~\cite{physics_based_detection}, machine-learning techniques~\cite{machine_learning_approach}, or access-control mechanisms~\cite{ac_abac1,ac_abac2}. 


In sum, the proposed work tackles the problem of \emph{quantifying the practical tolerability of enforcing a strong full-stack memory-safety 
on realistic 
CPS with hard real-time constraints and limited computational power.} Furthermore, \emph{this work tackles the problem of ensuring availability of critical services and systems 
while successfully detecting and 
mitigating a wide-range of memory-safety attacks.}


We make the following contributions:
\textbf{a)} 
The enforced full-stack memory-safety effectively prevents both 
runtime crashes (that could arise due to memory-safety violations) and memory-safety attacks in CPS, both in user-space and kernel-spaces.
\textbf{b)} We formally define and 
model the notions of real-time and physical-state resiliency constraints, that are crucial in the context of CPS. 
\textbf{c)} We empirically measure and quantify tolerability of the induced performance overhead of our full-stack memory-safety 
based on the real-time constraints of two 
realistic CPS systems.
\revise{
\textbf{d)} Our 
user-space memory-safety ensures system availability and physical-state resiliency with reasonable 
performance and storage overheads.} Therefore, it is practically applicable to systems with stringent timing constraints, such as CPS, beyond the mainstream systems. 
\textbf{e)} The efficiency and effectiveness of our approach is evaluated on two real-world CPS testbeds containing vendor-supplied PLCs. 

\section{Attacker and system models}\label{attacker_and_system_model}
This section discusses our attacker model and the CPS design constraints we formally modelled.   

\subsection{Attacker model}\label{attacker_model}
\begin{figure}[t]
\centering
    \subfloat[Memory layout with relevant addresses]{
        \includegraphics[width=0.45\textwidth]{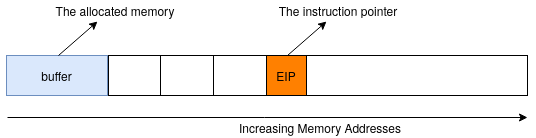}
        \label{memory_layout}
  }
    \hfil
    \subfloat[The attacker created tailored input]{
        \includegraphics[width=0.45\textwidth]{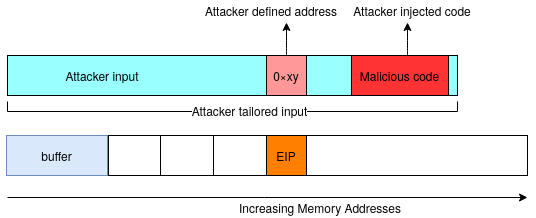}
        \label{tailored_input}
    
    }
    \hfil
    \subfloat[Diverting control to the injected code (code injection attacks) or system modules (code reuse attacks)]{
        \includegraphics[width=0.45\textwidth]{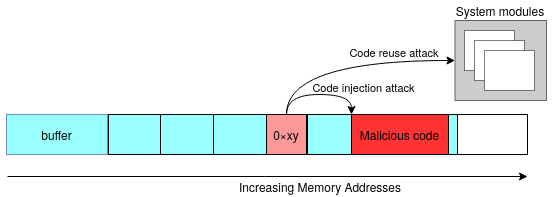}
        
         \label{memory_safety_attacks}
    }
    \label{memory_safety_exploits}
  \caption{A high-level illustration of memory-safety attacks exploitation strategies}
\end{figure}

The main 
objective of memory-safety attacks (e.g. code-injection and code-reuse attacks) is to get a privileged access or to take control 
of the vulnerable system. To achieve this, the attacker exploits memory-safety vulnerabilities, e.g., buffer over/under-flows and dangling pointers, that can be found in the targeted program. 
We briefly illustrate the exploitation strategy using a simple C/C++ program consisting of a buffer overflow vulnerability (cf. Program \ref{buffer_overflow_example}). 
\revise{A relevant memory layout of the program is 
provided in Figure~\ref{memory_layout}, 
just to 
simplify the illustration of the exploitation strategy. This includes the defined buffer address and extended instruction pointer (EIP) of the program.} 

\begin{figure}[htb]
 \centering
 \includegraphics[scale=0.38]{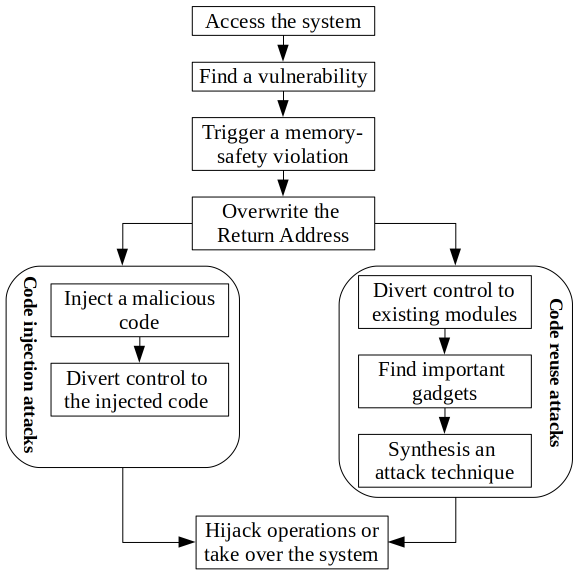}
 \caption{Memory-safety exploitation strategies}
 \label{fig:attacker_model}
\end{figure}

The vulnerable function, i.e., {\em ``gets(buffer)''}, 
allows the attacker to send an input data that is larger than the allocated buffer size. 
The attacker can exploit this vulnerability by creating a systematically tailored input that serves to overwrite the buffer's boundary, the EIP and other important memory addresses. In brief, the tailored input consists of the {\em attacker defined memory address} (e.g. 0{\scriptsize $\times$}xy in Figure \ref{tailored_input}) -- which will serve for overwriting the EIP, and a {\em malicious code} -- 
to be injected into the program's address space (in case of code-injection attacks). 
Figure \ref{tailored_input} illustrates the tailored input. As shown in Figure \ref{memory_safety_attacks}, the attacker defined address 
is made to point the starting address of the injected malicious code (for code-injection attacks) or existing system modules (for code-reuse attacks). 
The attack will be then launched by sending the 
tailored input to the buffer. 
The exploitation strategies are briefly illustrated in Figure \ref{fig:attacker_model}. A detailed account of such exploitation strategies can also be found in \cite{brop,coop}.



\begin{cppalgorithm}[htb]
\small 
\SetKwProg{Fn}{}{ \{}{\}}
\Fn(){foo()}{
	char buffer[16];\\
    printf(``Insert input: '');\\
    {\bf gets(buffer)};\label{gets_fun}\\
}
\caption{A code snippet containing a simple buffer overflow vulnerability}
\label{buffer_overflow_example}
\end{cppalgorithm}

\subsection{Modeling CPS design constraints}\label{modeling_cps_design_constraints}

\begin{figure}[htb]
 \centering
 \includegraphics[scale=0.4]{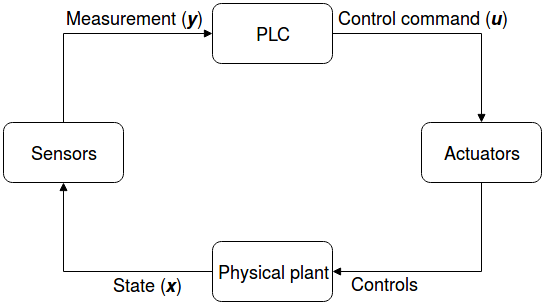}
 \caption{An abstraction of a typical CPS}
 [{\bf Acronyms:} $x$ = state vector, $y$ = sensor measurements, $u$ = control command]
 \label{fig:cps_model}
\end{figure}

Unlike traditional IT systems, CPS involves complex and continuous 
interactions between entities in the physical and cyber spaces over communication networks. These 
interactions are accomplished 
via communications with the physical-world through sensors and actuators and with the digital-world through PLCs (controllers) and other embedded devices. 
An abstraction of a typical 
CPS 
is illustrated in Figure~\ref{fig:cps_model}. 

The interactions among CPS entities, such as sensors, PLCs and actuators, is 
synchronized via system time. 
These interactions, unlike in conventional IT systems, are 
constrained by hard deadlines. Missing deadlines could 
result in disruption of the control-loop stability or damage to the physical plant (in the worst case). Because, such situations could lead the underlying system to run into an unsafe and unstable states. This is the main reason that makes CPS to be highly delay sensitive and real-time constrained systems. 


In particular, since PLCs form the main control devices in CPS, the real-time requirements are particularly imposed in these devices to maintain the safety and stability of the control system in CPS. 
To be able to formally capture the timing constraints in CPS, we define two crucial notions, namely \emph{real-time constraints} and \emph{physical-state resiliency}, 
which will serve as 
metrics to evaluate the efficiency and resilience of our full-stack memory-safety enforcement, respectively. We formally define and discuss these notions in the following sections. 



\subsubsection{\textbf{Real-time constraints}}\label{modeling_real_time_constraints}

\begin{figure}[htb]	
	\centering
        \includegraphics[scale=0.43]{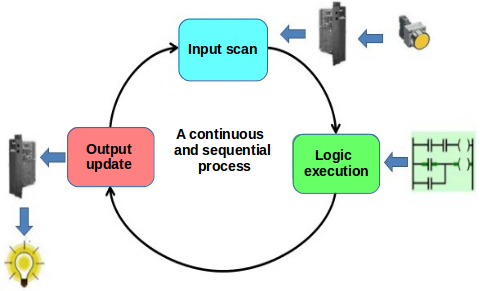}
  		\caption{The scan cycle of a PLC}
  		\label{fig:scan_cycle}
\end{figure}

\revise{
As shown in Figure \ref{fig:scan_cycle}, PLCs undergo a continuous and cyclic process 
when issuing control commands to actuators. This process involves three main operations, namely input scan, PLC logic execution and output update. This cyclic process is often referred as the PLC's \emph{scan cycle}. }
The overall time it takes to complete the scan cycle (i.e. to execute the three operations) is referred as the \emph{scan time} ($T_{s}$) of the PLC~\cite{ics-sea}. 
\revise{
To effectively 
synchronize the interactions and communications among its various entities, a typical CPS defines an upper-bound scan time to each PLC, called \emph{cycle time} ($T_c$).} Meaning, each scan cycle has to be completed within the specified cycle time of the PLC, i.e., $T_{s} \leq T_c$. We 
define this requirement as the \emph{real-time constraint} of the PLC. A typical PLC meets this constraint by design. However, due to security overheads, such as 
MSO, PLCs might not meet this constraint. \revise{
For example, by hardening the PLC with our memory-safety protection, the scan time increases. This increase in the scan time is attributed to the MSO. Concretely, the  
MSO can be computed as follows:}
\begin{equation}\label{eq:mso}
\textit{MSO} = \hat{T}_{s} - T_{s},
\end{equation}
where $\hat{T}_s$ and $T_{s}$ are the scan time with and without memory-safe compilation, respectively. 

\revise{The induced MSO by the memory-safe compilation obviously causes a delay on the PLC operations. However, it is essential to check whether 
this MSO still satisfies the real-time constraint imposed by the PLC. To this end, we compute MSO for -- 1) average-case and 2) worst-case scenarios. 
In brief, after 
securely compiling the PLC with our memory-safety, we measure the PLC scan time, i.e. $\hat{T}_{s}$, for $n$ different 
scan cycles. Then, we compute the 
MSO in average-case (i.e. $mean(\hat{T}_{s})$) and worst-case (i.e. $max(\hat{T}_{s})$) scenarios. 
Formally, we say that the MSO is acceptable 
in average-case 
if the following condition is satisfied: 
}
\begin{equation}
\label{eq:average_case_tolerability}
\frac{\sum_{i = 1}^{n} \hat{T}_s(i)}{n} \leq T_c 
\end{equation}
where $\hat{T}_{s}(i)$ captures 
the scan time for the $i$-th measurement after the memory-safe compilation. 

\revise{
Similarly, the MSO is acceptable 
in the worst-case if the following condition is satisfied: 
}
\begin{equation}
\label{eq:worst_case_tolerability}
\displaystyle \max_{i = 1}^{n}\ \hat{T}_s(i) \leq T_c 
\end{equation}

\revise{
Note that since cyber-physical systems are hard real-time constrained systems, each scan time should meet the PLC's real-time requirement. As such, the worst-case scenario should be used 
to ensure real-time guarantees in CPS. However, the worst-case MSO (i.e. the highest cycle time obtained out of $n$ scan cycles) might not reflect the actual overhead of the enforced memory-safety. This is because, some scan times could be inflated even due to non-MSO related reasons, e.g. sudden execution interruptions due to unforeseen reasons. For this reason, it is essential to demonstrate the average MSO as well 
since it gives an intuition 
of the average performance penalties to be paid 
when enforcing this memory-safety solution even in other CPS testbeds. Therefore, we demonstrate both the average-case (computed out of 50,000 scan cycles) and worst-case MSO in this paper. 
}

\revise{
\subsubsection{\textbf{Physical-state resiliency}}\label{modeling_physical_state_resiliency}

As discussed in the preceding sections, the stability of PLCs is crucial in enforcing the dynamics of a CPS to be compliant with its requirements. The PLCs real-time constraints, 
discussed in Section~\ref{modeling_real_time_constraints}, play a crucial role 
to stabilize the PLC operations as well as 
to properly synchronize the interactions among various entities in CPS. However, these 
constraints are mainly for the cyber-world (where controls and communications take place) and might not clearly 
reflect the impact of control delays 
on the physical-world (where physical processes take place). In this section, we particularly model the impact of control delays (or PLC downtime, in other words) on the physical 
plant in CPS. For example, 
a PLC issues a control command 
at the rate of its cycle time (i.e. $T_c$) and the respective actuator also receives this command at the same rate. 
If 
a control delay happens for an arbitrary amount of time, say $\tau$, then the actuator will not receive a fresh control command for the duration of $\tau$. Consequently, the physical dynamics of the 
CPS will be affected for a total of $\frac{\tau}{T_c}$ scan cycles. 

We note that the duration $\tau$ might be arbitrarily large depending on the 
reason that causes the control delay. For example, the 
control delay could happen because of the MSO (cf. Section \ref{modeling_real_time_constraints}) or as a result of a non-resilient mitigation strategy that typically 
aborts/restarts the PLC (i.e. causing PLC downtime) when a memory-safety attack is detected (cf. Section \ref{asan_validation}). 

In either case, the scan time of the PLC with the enforced memory-safety (i.e. $\hat{T}_s$) may increase beyond the cycle time (i.e. $T_c$). This may affect 
the dynamics of the physical processes in CPS. In the worst case, this 
delay could cause 
damage to the 
CPS by violating its upper-bound or lower-bound physical state (i.e. $x$) limits of the plant. 

For example, let us take the first process in SWaT (discussed in Section \ref{swat}). 
This process controls the inflow of water from an external water supply to a raw water tank. PLC1 controls this process by opening (with ``ON'' command) and closing (with ``OFF'' command) a motorized valve, i.e., the actuator, connected with the inlet pipe to the tank. If the valve is ``ON'' for an arbitrarily long duration, then the raw water tank overflows when the water level surpasses the upper-bound limit of the tank. 
This occurs due to the control delay $\tau$ on PLC1, during which, the control command (i.e. ``ON'') computed by PLC1 may not change. Similarly, if the actuator receives the ``OFF'' command from PLC1 for an arbitrarily long duration, then the water tank underflows when the water level goes beyond the lower-bound limit of the tank. This is because tanks from other processes 
expect raw water from this underflow tank. The occurrence of such phenomena could severely affect the system dynamics in CPS. 


Here, we quantify the {\em tolerability} of the 
control delay $\tau$, i.e., the length of $\tau$ that 
does not violate the upper-bound and lower-bound physical state limits of the plant. We define this notion of tolerance as \emph{physical-state resiliency}.  

The tolerability of $\tau$, 
in fact, depends on the current physical-state of the plant (e.g. water level, in case of PLC1 in SWaT) and the last control command issued by the PLC (e.g. "ON" or "OFF" command). In the following, we will formally define $\tau$ and the notion of \emph{physical-state resiliency} in CPS. 

 To accurately formulate the control delay $\tau$, we need to consider the following three mutually exclusive scenarios:
\begin{enumerate}
\item The 
PLC is aborted or restarted.
\item The 
PLC is neither aborted nor restarted and $\hat{T}_s \leq T_c$. In this case, there will be no observable impact on the physical dynamics of the 
CPS. This is because the PLCs, despite having increased scan time, still meet the real-time constraint $T_c$. Thus, they are not susceptible to 
control delays.
\item The 
PLC is neither aborted nor restarted and $\hat{T}_s > T_c$. In this case, the PLCs will have a 
control delay of $\hat{T}_s - T_c$, as the scan time violates the real-time constraint $T_c$.
\end {enumerate}

Based on the intuitions 
discussed in the preceding paragraphs, we formally define $\tau$ as follows:
\begin{equation}
\label{eq:downtime}
\tau = \begin{cases}
\Delta, &\text{PLC is aborted/restarted}\\
0, &\text{$\hat{T}_s \leq T_c$}\\
\hat{T}_s - T_c, &\text{$\hat{T}_s > T_c$}
\end{cases}
\end{equation}
where $\Delta$ captures a non-deterministic threshold on the 
control delays when the PLC is aborted or restarted.  

To formally model the physical-state resiliency, we will take a control-theoretic approach. For the sake of simplicity, we will assume that the dynamics of a typical CPS, without considering the noise and disturbance on the controller, is modeled via a linear-time invariant. This is formally captured as follows (cf. Figure~\ref{fig:cps_model}): 
\begin{equation}
\label{eq:modeling_state}
x_{t+1} = Ax_{t} + Bu_{t}
\end{equation}
\begin{equation}
\label{eq:modeling_output}
y_{t} = Cx_{t}
\end{equation}
where $t \in \mathbb{N}$ captures the index of discretized time domain. $x_t \in \mathbb{R}^{k}$ is the state vector of the physical plant at time $t$, $u_t \in \mathbb{R}^{m}$ is the control command vector at time $t$ and $y_t \in \mathbb{R}^{k}$ is the measured output vector from sensors at time $t$. $A \in \mathbb{R}^{k \times k}$ is the state matrix, $B \in \mathbb{R}^{k \times m}$ is the control matrix and $C \in \mathbb{R}^{k \times k}$ is the output matrix. 

We now consider a duration $\tau \in \mathbb{R}$ for the 
control delay. With the 
control delay $\tau$, we revisit Eq.~(\ref{eq:modeling_state}) and the state estimation is refined as follows: 
\begin{equation}
\label{eq:modeling_state_in_delay}
x'_{t+1} = Ax_{t} + Bu_{t-1}[\![ t, t+\tau ]\!]
\end{equation}
where $x'_{t+1} \in \mathbb{R}^{k}$ is the estimated state vector at time $t+1$ and 
there was a control delay for a maximum duration $\tau$. The notation $u_{t-1}[\![ t, t+\tau ]\!]$ captures that the control command $u_{t-1}$ was active for a time interval $[t, t+\tau]$ due to 
the control delay $\tau$. In Eq.~(\ref{eq:modeling_state_in_delay}), we assume, without loss of generality, that $u_{t-1}$ is the last control command received from the PLC before 
the occurrence of the control delay. 

To check the tolerance of $\tau$, we need to validate the physical state vector $x_t$ at any discretized time index $t$. To this end, we first assume an upper-bound $\omega \in \mathbb{R}^{k}$ and lower-bound $\theta \in \mathbb{R}^{k}$ thresholds on the physical state vector $x_t$. 
Therefore, to satisfy the physical-state resiliency, $x_t$ must not exceed 
$\omega$ nor subceed $\theta$. 
Formally, we say that a typical CPS (cf. Figure~\ref{fig:cps_model}) satisfies physical-state resiliency if and only if the following condition holds at an arbitrary time index $t$: 
\begin{equation*}
\theta \leq  x'_{t+1} \leq \omega
\end{equation*}
\begin{equation}
\label{eq:psr}
\theta \leq Ax_{t} + Bu_{t-1}[\![ t, t+\tau ]\!] \leq \omega 
\end{equation}

Figure~\ref{fig:downtime} illustrates three representative scenarios to show the consequence of Eq.~(\ref{eq:psr}). If the  
control delay $\tau_1 =0$, then $u_t$ (i.e. control command at time $t$) is correctly computed and $x'_{t+1}=x_{t+1}$. If the 
control delay $\tau_2 \in (1,2]$, then the control command $u_{t}$ will be the same as $u_{t-1}$. Consequently, $x'_{t+1}$ is unlikely to be equal to $x_{t+1}$. Finally, when 
the control delay $\tau_3 > 2$, the control command vector $u_{t+i}$ for $i \ge 0$ will be the same as $u_{t-1}$. As a result, the estimated state vectors $x'_{t+j}$ for $j \ge 1$ will unlikely to be identical to $x_{t+j}$. 
}

\begin{figure}[htb]
 \centering
 \includegraphics[scale=0.3]{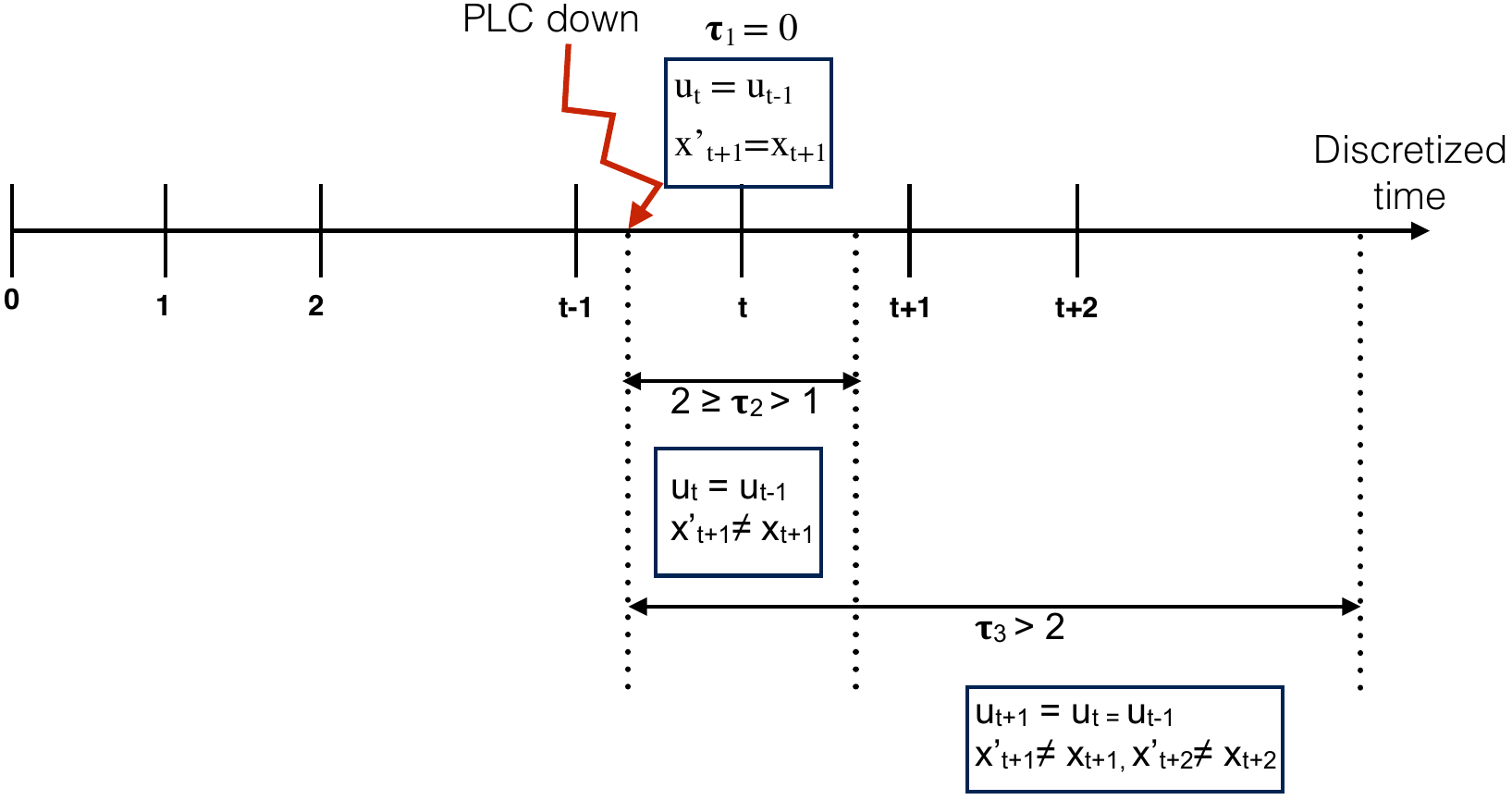}
 \caption{\revise{Illustrating the impact of control delays in CPS}} 
 \label{fig:downtime}
\end{figure}

\section{Secure compiling of PLCs}\label{memory_safety_tools}
%
In this section, we present a high-level discussion of the memory-safety tools we used in our full-stack memory-safety enforcement. 

\subsection{ASan: user-space detection}\label{asan}

\subsubsection{Overview}

Although several memory-safety tools are available, 
they might not be practically applicable in CPS because of various reasons. This includes limitations in error coverage, low detection accuracy, high performance overheads, ineffective mitigation strategy, and architectural incompatibilities, among others. After researching over various tools, we chose ASan~\cite{asan} as our user-space memory error and attack detector tool 
because of its high detection accuracy, broader error coverage and relatively low performance overhead when compared with other \revise{code-instrumentation based memory-safety} tools~\cite{asan,asgithub}.



 ASan instruments C/C++ programs at compile-time and creates poisoned memory regions, known as {\it redzones}. These redzones are not addressable and any instruction attempting to access them 
 will be proactively detected as a memory safety violation. The instrumentation of ASan is illustrated in Figure~\ref{fig:cfg_scenario1} and \ref{fig:cfg_scenario2}. \revise{Such an instrumented code can detect 
 numerous memory-safety vulnerabilities, such as buffer overflows (i.e. stack, heap and global buffer overflows), use-after-free
errors (dangling pointers), use-after-return errors, initialization order bugs, memory leaks, and double free errors.}

\subsubsection{Porting ASan to CPS}

Since ASan is specifically designed for the x86 target architectures, 
it is incompatible with AVR or ARM based architectures in the CPS environment. In this work, we adapt ASan for ARM based PLC systems and evaluate its applicability in the context of cyber-physical systems. PLC instructions as well as firmware of the PLC are compiled and instrumented by a modified ASan to guarantee a notion of memory-safety when the program is executed at runtime.

\subsubsection{Validation}\label{asan_validation}

The main limitation of ASan is its 
non-resilient mitigation strategy; it simply aborts the victim program whenever a 
memory violation or an attack is detected (cf. Figure~\ref{fig:cfg_scenario1} and \ref{fig:cfg_scenario2}). Ultimately,  
cyber-physical systems require a notion of security that guarantees correct functioning of the control system under strong attacker models. However, ASan  fails at that as the detection of a memory-safety violation leads to abortion of the PLC program and may leave the control system in an unsafe state. This makes ASan inapplicable in systems with stringent availability constraints, such as CPS. 

To address this mitigation limitation, we recently implemented 
CIMA~\cite{cima_cose} -- a resilient mitigation technique against memory-safety attacks -- and integrated it with ASan (see Section \ref{cima}).  
Thus, CIMA enhances the capability of ASan to mitigate memory-safety bugs on-the-fly. 

\subsection{CIMA: user-space mitigation}\label{cima}

\subsubsection{Overview}\label{cima_overview}

CIMA~\cite{cima_cose} 
is a resilient mitigation strategy we recently implemented to 
address the mitigation limitation of ASan. Instead of aborting the victim program upon detection of a memory-safety violation, CIMA proactively counters memory-safety violations from occurring. This is accomplished by 
proactively skipping (i.e. not executing) the illegal memory access instructions , i.e., instructions that attempt to access memory illegally, at runtime. 
In such a way, CIMA effectively prevents memory-safety violation from occurring without aborting/restarting the victim system. This  preserves system availability and 
also maintains physical-state resiliency in CPS even in the presence of memory-safety attacks. 

\subsubsection{Detailed methodology}\label{cima_methodology}

\begin{figure*}[t]
 \centering
 \includegraphics[scale=0.47]{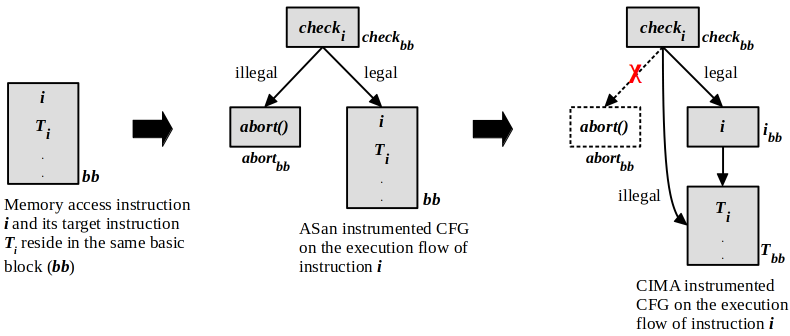}
 \caption{Construction of the CFG 
 when $i$ and $T_{i}$ reside in the same basic-block
 }
 \label{fig:cfg_scenario1}
\end{figure*}

\begin{figure*}[t]
 \centering
 \includegraphics[scale=0.47]{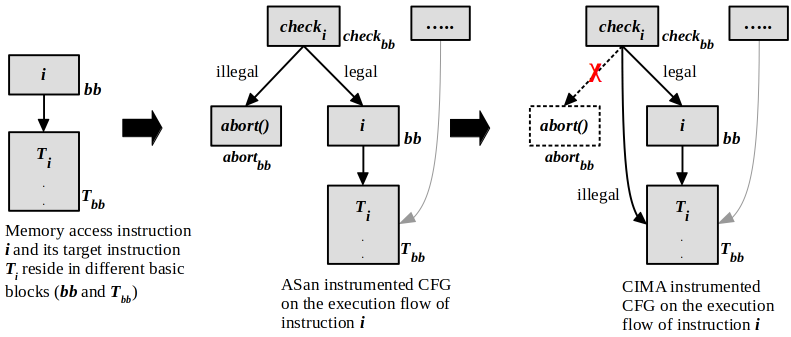}
 \caption{Construction of the CFG 
 when $i$ and $T_{i}$ reside in different basic-blocks
 }
 \label{fig:cfg_scenario2}
\end{figure*}

To bypass 
illegal memory accesses, 
CIMA systematically constructs and manipulates the compiler-generated control-flow graph (CFG) of the program. CIMA constructs the CFG at compile-time by instrumenting each memory access instruction of the program. The instrumentation involves computation of the target instruction $T_{i}$ for each respective memory access instruction $i$. The target instruction $T_{i}$ is computed 
as a single successor of the memory access 
instruction $i$ in the CFG, which will be determined at runtime. 
So, the newly constructed CFG will contain each memory access instruction $i$ and its 
corresponding target instruction $T_{i}$ 
(cf. Figure \ref{fig:cfg_scenario1} and \ref{fig:cfg_scenario2}). In such a fashion, if 
instruction $i$ is detected as illegal at runtime, $i$ will be then bypassed and 
its target instruction $T_i$ is executed instead, hence preventing the execution of the illegal memory access instruction. If $T_{i}$ is also detected as illegal, 
the successor of $T_{i}$ will be then executed, and so on. The rest of the execution, nevertheless, continues without interruption. This ensures availability of the program even in the presence of memory-safety attacks or violations.

Furthermore, it is noteworthy that the construction of the new CFG involves two scenarios depending on the location of $i$ and $T_{i}$ in the original CFG. 

\noindent {\it Scenario 1}: When the memory access instruction $i$ and its target instruction $T_{i}$ are resided in the same basic-block (say $bb$). In this case, it is not possible to make a conditional jump to $T_{i}$ within the same basic-block $bb$ if $i$ is detected as illegal. 
Thus, to make the conditional jump possible, the basic block $bb$ is split to two basic blocks -- $i_{bb}$ (containing $i$) and $T_{bb}$ (containing $T_{i}$). Now, control can jump to $T_{bb}$ if instruction $i$ is detected as illegal. The newly constructed CFG in this scenario is illustrated in Figure \ref{fig:cfg_scenario1}. 

\noindent {\it Scenario 2}: When $i$ and $T_{i}$ are resided in different basic-blocks. In this case, 
we do not need to split the basic-block since $i$ and $T_{i}$ are resided in their respective basic-blocks in the original CFG. So, control can simply jump to the target basic-block (i.e. $T_{bb}$) when instruction $i$ is detected as illegal. The CFG construction of this scenario is depicted in Figure \ref{fig:cfg_scenario2}. 

A more discussion of the two scenarios as well as a detailed account of CIMA can be found in \cite{cima_cose,cima_arxiv}.

\subsubsection{Validation}\label{cima_validation}

Although CIMA effectively mitigates memory-safety attacks, there exists semantics-preserving issues as skipping instructions affects the original program semantics. A careful discussion of such issue is provided in the original paper~\cite{cima_cose}. 
From the discussion, there are only a few corner cases (which might not even happen in a real-world) where CIMA might not be effective (see in\cite{cima_cose}). Apart from that, CIMA does not affect the execution of the program and this is validated experimentally. 
A detailed discussion of CIMA's validation can be found in \cite{cima_cose}.

\subsection{KASan: kernel-space detection}\label{kasan}

\subsubsection{Overview}

Nowadays, most PLCs are user-mode applications running on POSIX-like OSs, such as Linux OS \cite{eyasu_essos}. For instance, Allen Bradley PLC5 
uses \emph{Microware OS-9} \cite{tofino_security}; 
Siemens SIMATIC\cite{siemens_simatic} uses \emph{Microsoft Windows}; Schneider Quantum uses \emph{VxWorks} \cite{tofino_security}; Emerson DeltaV uses \emph{VxWorks} \cite{tofino_security}; 
and OpenPLC uses \emph{Linux OS} \cite{openplc}. 

Therefore, in addition to the exploitation threats at user-space, 
cyberattacks may also exploit memory-safety vulnerabilities that could be found in the underlying operating systems hosting the PLCs. 
In particular, attacks may exceptionally target vulnerabilities in the Linux kernel (as recent trends also show in CVE \cite{cve_linux}). 
Therefore, the underlying OS kernel of the PLC is another attack surface for memory-safety attacks targeting CPS. 


To address the kernel-space security concern, we ported KASan in CPS 
by fixing its various architectural incompatibility issues. KASan \cite{kasan} 
is a fast and dynamic memory error detector tool 
designed for the Linux kernel. 

KASan is also a code-instrumentation based tool and it follows a similar approach with ASan to detect memory-safety violations. 
However, with the assumption of not to heavily affect performance of the Linux kernel, KASan is 
made to cover only a limited (but critical) set of memory-safety vulnerabilities, such as buffer 
overflows, double-free errors and use-after-free bugs. 
Consequently, its runtime overhead is considerably lower when compared with that of ASan.

KASan also follows a similar mitigation strategy with that of ASan. It automatically aborts the victim program 
when a memory violation is detected, which is an ineffective and non-resilient mitigation strategy as discussed earlier. 

\subsubsection{Porting KASan in CPS}\label{enforcing_kasan}

Porting KASan to our CPS setup was not a straightforward task because of architectural incompatibility and other technical issues. 
The current version of KASan is 
designed only for the x86\_64 and ARM64 target architectures \cite{kasan,eyasu_essos}. Unfortunately, Raspbian (i.e. the underlying operating system of Raspberry PI) is a 32-bit OS, and no support for the 64-bit architecture so far. Hence, the Raspbian kernel (at the time of writing this paper) is based on a 32-bit (i.e. ARM32) target architecture. For this reason, it is not possible to directly enforce KASan to the Raspbian kernel. Besides, we also encountered several technical difficulties when the kernel-level building tool (i.e. GCC) is different from the user-level GCC (which incorporates ASan and CIMA). 

To overcome these technical problems, it was essential to build a custom Linux kernel with ARM64 architecture. To accomplish this, first we built a cross-compiler toolchain (using ASan and CIMA enabled GCC) on a 64-bit Linux OS. Then, we cross-compiled a custom Raspberry PI Linux kernel (with 64-bit architecture) using our cross-compiler toolchain. Finally, we managed to compile the custom Linux kernel with KASan (by enabling the \texttt{KASAN=y} compiler switch and other configuration flags). This hardened Linux kernel can then detect kernel-level memory-safety attacks or violations, such as buffer overflows, use-after-free bugs and double-free memory errors. 

Similar to the cases in ASan and CIMA, our experimental setup has also allowed us to empirically measure the performance overhead of KASan. Then, we quantify its performance impact and hence acceptability in the CPS context. 


\section{Experimental design}\label{experimental_design}  

The effectiveness of our proposed security measures against memory-safety attacks is experimented on realistic CPS testbeds. This section presents a brief discussion of the two CPS testbeds used to conduct our experiments. 

\subsection{SWaT}\label{swat} 

\begin{figure}[htb]
	\centering
  \includegraphics[scale=0.41]{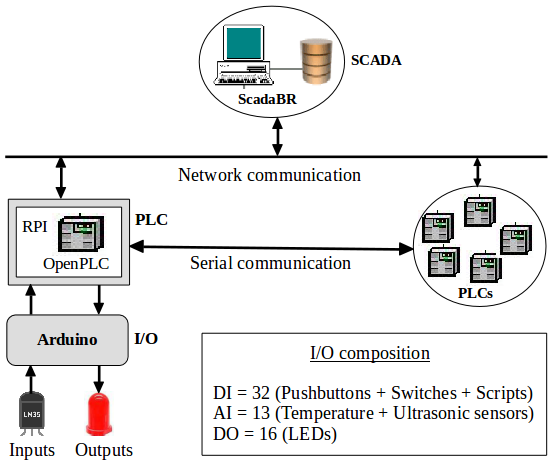} 
	\caption{Architecture of Open-SWaT}
	\label{fig:openswat}
\end{figure}

\revise{
SWaT~\cite{swat} is a fully operational water purification testbed for research in the design of secure cyber-physical systems.
It produces five gallons/minute of doubly filtered water. In the following, we briefly discuss 
the purification process in SWaT and how we reproduce SWaT to conduct our experiments.  

\subsubsection{Purification process}\label{purification_process}

The entire water purification process is carried out by six distinct, yet co-operative, sub-processes. Each sub-process is controlled by an independent PLC (indexed from PLC1 through PLC6). Specifically, PLC1 controls the first sub-process, i.e., the inflow of water from external supply to a raw water tank, by opening and closing the motorized valve connected with the inlet pipe to the tank. PLC2 controls the chemical dosing process, e.g., water chlorination, where appropriate amount of chlorine and other chemicals are added to the raw water. PLC3 controls the ultrafiltration (UF) process. PLC4 controls the dechlorination process where any free chlorine is removed from the water before it is sent to the next stage. PLC5 controls the reverse osmosis (RO) process where the dechlorinated water is passed through a two-stage RO filtration unit. The filtered water from the RO unit is sent in the permeate tank, where the recycled water is stored, and the rejected water is sent to the UF backwash tank. In the final stage, PLC6 controls the cleaning of the membranes in the UF backwash tank by turning on and off the UF backwash pump. The overall purification process of SWaT is shown in Figure \ref{fig:purification_process}
The overall purification process of SWaT is shown in Figure \ref{fig:purification_process}. 
A detailed account of 
SWaT 
can be found in \cite{cima_cose,swat}. 

\begin{figure*}[htb]
	\centering
    \includegraphics[scale=0.29]{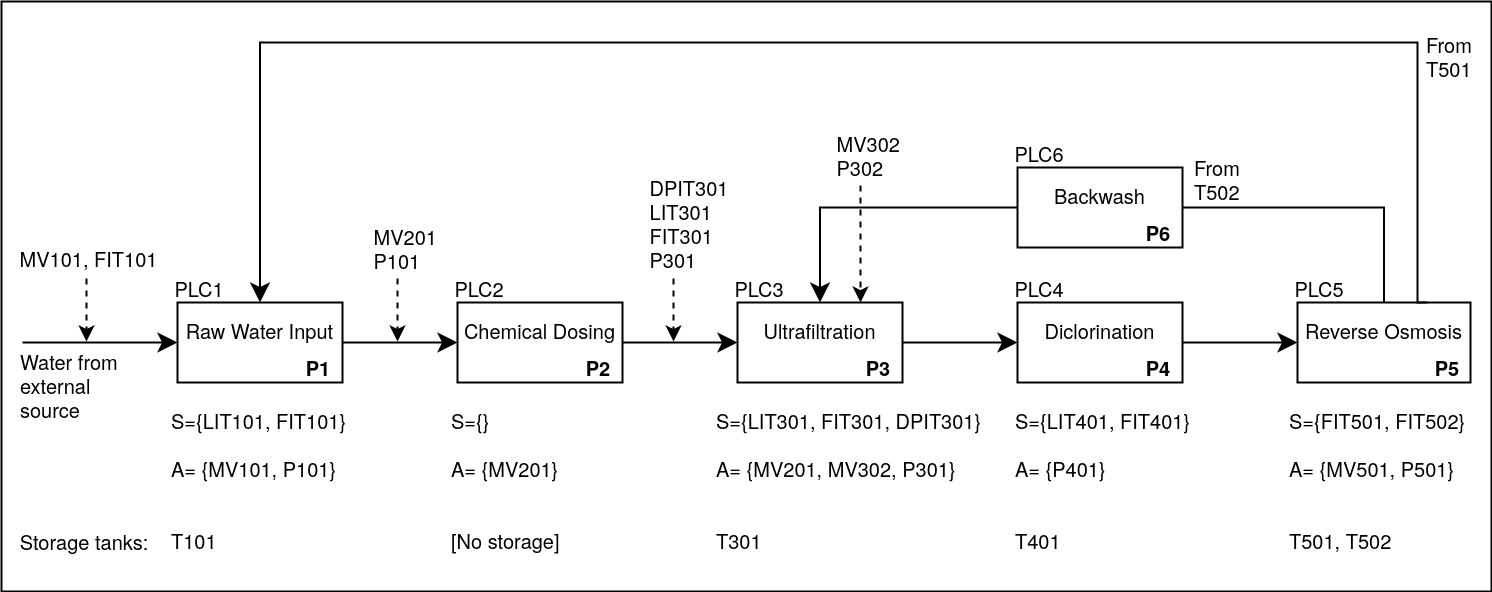}
    \caption{Overview of the water purification process in SWaT}
    [{\bf Definition of the acronyms:} S = Sensor, A = Actuator, T = Tank, P = Process, MV = Motorized Valve, LIT = Level Indicator Transmitter, FIT = Flow Indicator Transmitter, DPIT = Differential Pressure Indicator Transmitter]
	\label{fig:purification_process}
\end{figure*}

\subsubsection{Open-SWaT}\label{open_swat}

SWaT is designed using 
proprietary PLCs. Hence, it is not possible to directly enforce our memory-safety since we cannot modify the firmware of these PLCs. To address this problem, we designed an open testbed, named Open-SWaT. 

Open-SWaT~\cite{cima_cose,eyasu_essos} is a mini CPS we designed using open-source PLCs~\cite{openplc} 
by mimicking the features and operational 
details of SWaT.  
For example, it mimics the hardware specifications of the SWaT PLCs (e.g. a CPU speed of 200MHz and a user memory of 2MB), a Remote Input/Output (RIO) terminal (containing 32 digital inputs (DI), 16 digital outputs (DO)), 13 analog inputs (AI), the real-time constraints, the PLC program (containing 129 instructions), the communication frequencies and the full SCADA system. 
A high-level architecture of Open-SWaT is illustrated in Figure \ref{fig:openswat}. A detailed account of Open-SWaT can also be found in \cite{cima_cose}.
}

\subsection{SecUTS}\label{secuts}

SecUTS~\cite{secuts_paper} is a CPS testbed designed 
to secure a Metro SCADA 
system. 
A detailed account of SecUTS testbed can be found in \cite{secuts_paper,cima_cose}. 

Unfortunately, this testbed is also based on proprietary PLCs, hence we cannot directly enforce our full-stack memory-safety to these PLCs.  Consequently, we 
also prototyped Open-SecUTS (by mimicking the SecUTS testbed) using the OpenPLC controller. It 
comprises 6 DI (emergency and control buttons) and 9 DO (tunnel and station lightings, ventilation and alarms). Subsequently, we enforced our full-stack memory-safety to Open-SecUTS and evaluated its practical applicability in a Metro SCADA system. 

\section{Evaluation and discussion}\label{evaluation_and_discussion}
In this section, 
we evaluate and discuss the experimental results of our full-stack memory-safety enforcement in CPS. 
In particular, we evaluated the security guarantee (i.e. the detection and mitigation accuracy), the efficiency (i.e. tolerability of the overall overhead in CPS), 
the mitigation resilience and the memory usage overheads of the full-stack enforcement. 

\subsection{Security guarantees}\label{security_evaluation}


In this section, as a sanity check on our full-stack memory-safety enforcement, we evaluate the detection and mitigation accuracy of ASan, CIMA and KASan 
over various memory-safety vulnerabilities. 
As discussed in 
previous sections, ASan and KASan may rarely miss some memory violations, such as global buffer overflows and use-after-free bugs, if an attacker manages to 
corrupt regions outside the redzones. Due to this reason, ASan and KASan could give rare false positives on these violations. Apart from this, ASan and KASan effectively detects memory-safety violations. CIMA, on the other hand, accurately mitigates all the memory-safety violations detected by ASan, but only in user-space. Table \ref{table:accuracy_asan_cima_kasan} 
summarizes the detection and mitigation coverage of the three tools over various memory-safety vulnerabilities.   

\begin{table*}[htb!]
\centering
\caption{Detection and mitigation accuracy of the full-stack memory-safety}
\label{table:accuracy_asan_cima_kasan}
\resizebox{\textwidth}{!}{%
\begin{threeparttable}
\begin{tabular}{|l|l|l|l|l|l|l|}
\hline
\multirow{2}{*}{Vulnerabilities}                                         & \multicolumn{2}{c|}{ASan (Detection)} & \multicolumn{2}{c|}{CIMA (Mitigation)} & \multicolumn{2}{c|}{KASan (Detection)} \\ \cline{2-7} 
                                                                         & False positive    & False negative    & False positive     & False negative    & False positive     & False negative    \\ \hline
                                                                         \hline
Stack buffer overflow                                                    & No                & No                & No                 & No                & No                 & No                \\ \hline
Heap buffer overflow                                                     & No                & No                & No                 & No                & No                 & No                \\ \hline
Global buffer overflow                                                   & No                & Rarely\tnote{*}            & No                 & No                & No                 & No                \\ \hline
Use-after-free bugs                                                      & No                & Rarely\tnote{*}            & No                 & No                & No                 & No                \\ \hline
Use-after-return bugs                                                    & No                & No                & No                 & No                & Uncovered          & Uncovered         \\ \hline
Initialization order bugs                                                & No                & No                & No                 & No                & Uncovered          & Uncovered         \\ \hline
Memory leaks                                                             & No                & No                & No                 & No                & Partially          & Partially         \\ \hline
Double-free errors                                                       & No                & No                & No                 & No                & No                 & No                \\ \hline
\begin{tabular}[c]{@{}l@{}}Uninitialized memory \\ reads\end{tabular} & Uncovered         & Uncovered         & Uncovered          & Uncovered         & Uncovered          & Uncovered         \\ \hline
\end{tabular}
\begin{tablenotes}
  \item[*] The reasons are discussed in Section \ref{security_evaluation} and further details can also be found in the original paper~\cite{asan}.
  \end{tablenotes}
  \end{threeparttable}
}
\end{table*}





\subsection{Performance}\label{performance_evaluation}


In this section, we discuss the practical tolerability of the overall full-stack memory-safety overhead, i.e., ASan + CIMA + KASan ($\hat{T''_{s}}$), in CPS. A detailed performance report of our full-stack memory-safety enforcement, including the overhead contributed by each tool, is depicted in Table \ref{table:fs_swat_mso} (for Open-SWaT) and Table \ref{table:fs_secuts_mso} (for Open-SecUTS). According to the results, $\hat{T''_{s}}$ is 91.02\% (for Open-SWaT) and 85.49\% (for Open-SecUTS).

Subsequently, we evaluate tolerability of the full-stack memory-safety overhead both in the average-case and worst-case scenarios. Essentially, we have checked if the overall overhead satisfies the conditions defined on Eq.~(\ref{eq:average_case_tolerability}) (for average-case scenario) and Eq.~(\ref{eq:worst_case_tolerability}) (for worst-case scenario). 

\begin{figure*}[!h]
	\centering
	\subfloat[\revise{The average-case MSO for Open-SWaT}]{
    \includegraphics[scale=0.33]{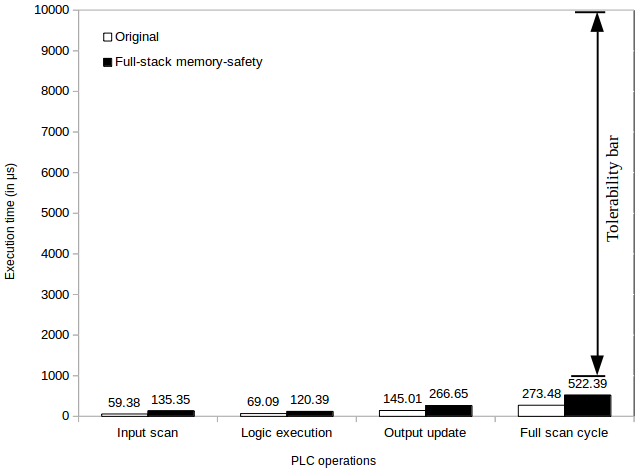} 
	\label{fig:chart_fullstack_swat_mean}
    } 
    \hfil
    \subfloat[\revise{The worst-case MSO for Open-SWaT}]{
    \includegraphics[scale=0.33]{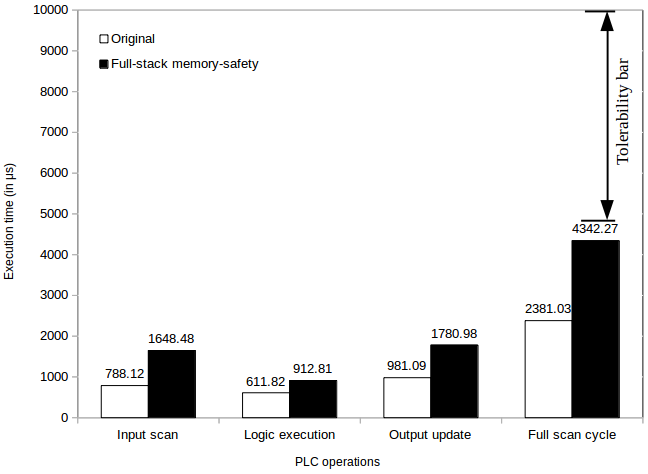} 
	\label{fig:chart_fullstack_swat_max}
    }
    \hfil
    \vspace{0.6cm}
    \subfloat[\revise{The average-case MSO for Open-SecUTS}]{
    \includegraphics[scale=0.33]{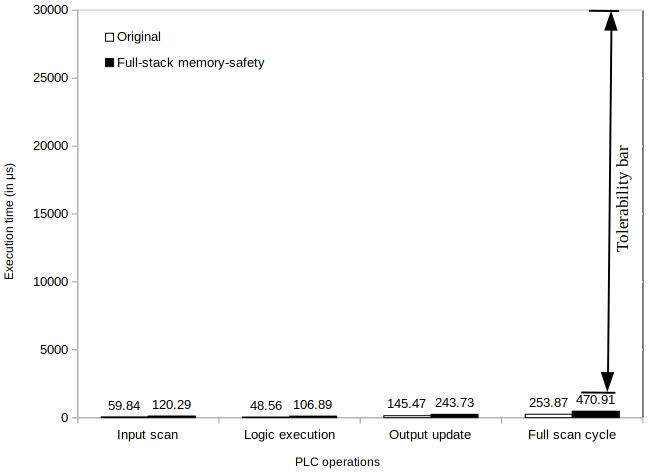}
	\label{fig:chart_fullstack_secuts_mean}
    } 
    \hfil
    \subfloat[\revise{The worst-case MSO for Open-SecUTS}]{
    \includegraphics[scale=0.33]{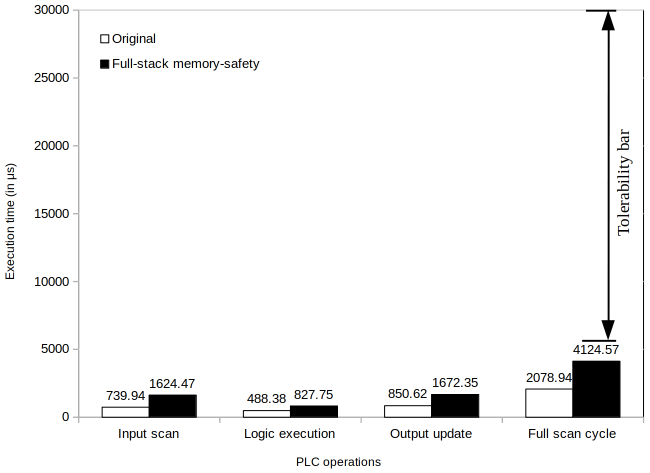}
	\label{fig:chart_fullstack_secuts_max}
    }
\caption{\revise{Tolerability of the full-stack MSO for the Open-SWaT and Open-SecUTS testbeds}}
\end{figure*}

First, we evaluated tolerability of the overhead in the average-case scenario. For Open-SWaT, $mean(\hat{T''_{s}}) = 522.39$µs (cf. Table \ref{table:fs_swat_mso}), and $T_{c} = 10000$µs; and for Open-SecUTS, $mean(\hat{T''_{s}}) = 470.91$µs (cf. Table \ref{table:fs_secuts_mso}), and $T_{c} = 30000$µs. Therefore, according to Eq.~(\ref{eq:average_case_tolerability}), the overhead is tolerable with a large magnitude for both SWaT and SecUTS in the average-case scenario (see the tolerability bar in Figure \ref{fig:chart_fullstack_swat_mean} and \ref{fig:chart_fullstack_secuts_mean}, respectively). 

Similarly, we evaluated tolerability of the full-stack overhead in the worst-case scenario. For Open-SWaT, $max(\hat{T''_{s}}) = 4342.27$µs (cf. Table \ref{table:fs_swat_mso}), and $T_{c} = 10000$µs; and for Open-SecUTS, $max(\hat{T''_{s}}) = 4124.57$µs (cf. Table \ref{table:fs_secuts_mso}), and $T_{c} = 30000$µs. Both overheads satisfy Eq.~(\ref{eq:worst_case_tolerability}), hence the overall overhead is tolerable for both SWaT and SecUTS even in the worst-case scenario (see the tolerability bar in Figure \ref{fig:chart_fullstack_swat_max} and \ref{fig:chart_fullstack_secuts_max}, respectively). That means, the overhead of our full-stack memory-safety 
largely meets the real-time constraints of SWaT and SecUTS both in the average-case and worst-case scenarios, while significantly increasing its security. 

\begin{table*}[htb]
\centering
\caption{MSO of the full-stack memory-safety for the Open-SWaT Testbed}
\label{table:fs_swat_mso}
\resizebox{\textwidth}{!}{%
\begin{tabular}{|l|l|l|l|l|l|l|l|l|l|l|l|l|l|l|l|l|l|}
\hline
\multirow{2}{*}{Operations} & \multirow{2}{*}{\begin{tabular}[c]{@{}l@{}}Number \\ of cycles\end{tabular}} & \multirow{2}{*}{\begin{tabular}[c]{@{}l@{}}Nodes\\ \end{tabular}} & \multirow{2}{*}{\begin{tabular}[c]{@{}l@{}}CPU\\  MHz\end{tabular}} & \multicolumn{2}{c|}{Original ($T_{s}$)}                                                                        & \multicolumn{4}{c|}{ASan ($\hat{T_{s}}$)}                                                                                                                                                                                      & \multicolumn{4}{c|}{ASan + CIMA ($\hat{T'_{s}}$)}                                                                                                                                                                              & \multicolumn{4}{l|}{ASan + CIMA + KASan ($\hat{T''_{s}}$)}                                                                                                                                                                     \\ \cline{5-18} 
                            &                                                                              &                                                                            &                                                                               & \begin{tabular}[c]{@{}l@{}}Mean\\ (in µs)\end{tabular} & \begin{tabular}[c]{@{}l@{}}Max\\ (in µs)\end{tabular} & \begin{tabular}[c]{@{}l@{}}Mean\\ (in µs)\end{tabular} & \begin{tabular}[c]{@{}l@{}}Max\\ (in µs)\end{tabular} & \begin{tabular}[c]{@{}l@{}}MSO\\ (in µs)\end{tabular} & \begin{tabular}[c]{@{}l@{}}MSO\\ (in \%)\end{tabular} & \begin{tabular}[c]{@{}l@{}}Mean\\ (in µs)\end{tabular} & \begin{tabular}[c]{@{}l@{}}Max\\ (in µs)\end{tabular} & \begin{tabular}[c]{@{}l@{}}MSO\\ (in µs)\end{tabular} & \begin{tabular}[c]{@{}l@{}}MSO\\ (in \%)\end{tabular} & \begin{tabular}[c]{@{}l@{}}Mean\\ (in µs)\end{tabular} & \begin{tabular}[c]{@{}l@{}}Max\\ (in µs)\end{tabular} & \begin{tabular}[c]{@{}l@{}}MSO\\ (in µs)\end{tabular} & \begin{tabular}[c]{@{}l@{}}MSO\\ (in \%)\end{tabular} \\ \hline
                            \hline
Input scan                  & 50000                                                                        & 6                                                                          & 200                                                                           & 59.38                                                  & 788.12                                                & 118.44                                                 & 1132.32                                               & 59.06                                                 & 99.46                                                 & 122.86                                                 & 1151.35                                                & 63.48                                                 & 106.9                                                 & 135.35                                                 & 1648.48                                               & 75.97                                                 & 127.94                                                \\ \hline
Execution           & 50000                                                                        & 6                                                                          & 200                                                                           & 69.09                                                  & 611.82                                                & 115.88                                                 & 720.36                                                & 46.79                                                 & 67.72                                                 & 118.97                                                 & 802.18                                                & 49.88                                                 & 72.2                                                  & 120.39                                                 & 912.81                                                & 53.3                                                  & 74.25                                                 \\ \hline
Output                & 50000                                                                        & 6                                                                          & 200                                                                           & 145.01                                                  & 981.09                                                & 185.37                                                 & 1125.45                                               & 40.36                                                 & 27.83                                                 & 199.89                                                 & 1213.62                                               & 54.88                                                 & 37.85                                                 & 266.65                                                 & 1780.98                                               & 121.64                                                & 83.88                                                 \\ \hline
Total              & 50000                                                                        & 6                                                                          & 200                                                                           & 273.48                                                 & 2381.03                                               & 419.69                                                 & 2978.13                                               & 146.21                                                & 53.46                                                 & 441.72                                                 & 3167.15                                               & 168.24                                                & 61.52                                                 & 522.39                                                 & 4342.27                                               & 248.91                                                & 91.02                                                 \\ \hline
\end{tabular}
}
\end{table*}
\begin{table*}[htb]
\centering
\caption{MSO of the full-stack memory-safety for the Open-SecUTS Testbed}
\label{table:fs_secuts_mso}
\resizebox{\textwidth}{!}{%
\begin{tabular}{|l|l|l|l|l|l|l|l|l|l|l|l|l|l|l|l|l|l|}
\hline
\multirow{2}{*}{Operations} & \multirow{2}{*}{\begin{tabular}[c]{@{}l@{}}Number \\ of cycles\end{tabular}} & \multirow{2}{*}{\begin{tabular}[c]{@{}l@{}}Nodes\\ \end{tabular}} & \multirow{2}{*}{\begin{tabular}[c]{@{}l@{}}CPU\\ MHz\end{tabular}} & \multicolumn{2}{c|}{Original ($T_{s}$)}                                                                        & \multicolumn{4}{c|}{ASan ($\hat{T_{s}}$)}                                                                                                                                                                                      & \multicolumn{4}{c|}{ASan + CIMA ($\hat{T'_{s}}$)}                                                                                                                                                                              & \multicolumn{4}{l|}{ASan + CIMA + KASan ($\hat{T''_{s}}$)}                                                                                                                                                                     \\ \cline{5-18} 
                            &                                                                              &                                                                            &                                                                               & \begin{tabular}[c]{@{}l@{}}Mean\\ (in µs)\end{tabular} & \begin{tabular}[c]{@{}l@{}}Max\\ (in µs)\end{tabular} & \begin{tabular}[c]{@{}l@{}}Mean\\ (in µs)\end{tabular} & \begin{tabular}[c]{@{}l@{}}Max\\ (in µs)\end{tabular} & \begin{tabular}[c]{@{}l@{}}MSO\\ (in µs)\end{tabular} & \begin{tabular}[c]{@{}l@{}}MSO\\ (in \%)\end{tabular} & \begin{tabular}[c]{@{}l@{}}Mean\\ (in µs)\end{tabular} & \begin{tabular}[c]{@{}l@{}}Max\\ (in µs)\end{tabular} & \begin{tabular}[c]{@{}l@{}}MSO\\ (in µs)\end{tabular} & \begin{tabular}[c]{@{}l@{}}MSO\\ (in \%)\end{tabular} & \begin{tabular}[c]{@{}l@{}}Mean\\ (in µs)\end{tabular} & \begin{tabular}[c]{@{}l@{}}Max\\ (in µs)\end{tabular} & \begin{tabular}[c]{@{}l@{}}MSO\\ (in µs)\end{tabular} & \begin{tabular}[c]{@{}l@{}}MSO\\ (in \%)\end{tabular} \\ \hline
                            \hline
Input scan                  & 50000                                                                        & 1                                                                          & 200                                                                           & 59.84                                                  & 739.94                                                & 114.88                                                 & 902.01                                                & 55.04                                                 & 91.98                                                 & 115.07                                                 & 906.09                                                & 55.23                                                 & 92.3                                                  & 120.29                                                 & 1624.47                                               & 60.45                                                 & 101.02                                                \\ \hline
Execution             & 50000                                                                        & 1                                                                          & 200                                                                           & 48.56                                                  & 488.38                                                & 91.36                                                  & 443.61                                                & 42.8                                                  & 88.14                                                 & 104.41                                                 & 676.19                                                & 55.85                                                 & 115.01                                                & 106.89                                                 & 827.75                                                & 58.33                                                 & 120.12                                                \\ \hline
Output                & 50000                                                                        & 1                                                                          & 200                                                                           & 145.47                                                 & 850.62                                                & 175.59                                                 & 1045.34                                               & 30.12                                                 & 20.71                                                 & 178.91                                                 & 924.11                                                & 33.44                                                 & 22.99                                                 & 243.73                                                 & 1672.35                                               & 98.26                                                 & 67.55                                                 \\ \hline
Total              & 50000                                                                        & 1                                                                          & 200                                                                           & 253.87                                                 & 2078.94                                               & 381.83                                                 & 2390.96                                               & 127.96                                                & 50.4                                                  & 398.39                                                 & 2506.39                                               & 144.52                                                & 56.93                                                 & 470.91                                                 & 4124.57                                               & 217.04                                                & 85.49                                                 \\ \hline
\end{tabular}
}
\end{table*}

\revise{
\subsection{Resilience}

As discussed in Section \ref{asan}, ASan simply aborts the victim program when a memory-safety violation or an attack is detected. Hence, it does not satisfy the physical-state resiliency requirement since the control delay $\tau$ is indefinite. However, the introduction of CIMA overcomes this mitigation limitation of ASan. To assess the physical-state resiliency of CIMA, we need to check if the control delay $\tau$, caused by the MSO or the mitigation strategy, satisfies Eq. (\ref{eq:psr}). In the former case, we already showed in the preceding section that the overall MSO induced by our full-stack memory-safety is tolerable, i.e., $\hat{T}_s \leq T_c$. Hence, the induced MSO does not affect the physical-state resiliency. In the later case, since CIMA does not abort or restart the PLC when mitigating memory-safety violations or attacks, it does not render system unavailability. That means, the control delay $\tau$ caused by our mitigation strategy is zero, hence Eq. (\ref{eq:psr}) is satisfied. Therefore, the physical-state resiliency constraint is satisfied for our user-space memory-safety enforcement. 

However, as discussed in Section \ref{kasan}, there is no a resilient mitigation strategy for the kernel-space since KASan simply aborts the victim program upon detection of memory-safety attacks or violations. As discussed, such mitigation strategy leads to system unavailability and hence not acceptable in CPS. 
Therefore, addressing this problem is left as a future work. 
}

\subsection{Memory usage overheads}\label{kasan_memory_usage_overheads}

The memory usage overhead of our full-stack memory-safety is also measured and evaluated. Table \ref{table:kasan_swat_memory_usage} and \ref{table:kasan_secuts_memory_usage} summarize the virtual memory, real memory, binary size and shared library usages for the Open-SWaT and Open-SecUTS testbeds, respectively. 

We notice a significant increase in virtual memory usages (11.11$\times$ for Open-SWaT and 10.92$\times$ for Open-SecUTS) in our full-stack enforcement. This is mainly due to the large redzones created with $malloc$ as part of the ASan and KASan approaches.  However, the real memory overhead is only 1.4$\times$ (for Open-SWaT) and 1.21$\times$ (for Open-SecUTS). We believe these overheads are still acceptable considering that most PLCs nowadays come with a minimum of 1GB memory size. Moreover, this memory overhead is an acceptable tradeoff in the light of strong countermeasures provided by our full-stack memory-safety solution. Finally, we observe that majority of the memory-usage overhead is incurred by ASan while KASan and CIMA only introduce a minimal and negligible memory usage overheads, respectively. 

\begin{table*}[htb]
\centering
\caption{Memory usage overheads of the full-stack memory-safety for Open-SWaT}
\label{table:kasan_swat_memory_usage}
\resizebox{\textwidth}{!}{%
\begin{tabular}{|l|l|c|c|c|c|l|l|}
\hline
\multirow{2}{*}{Category} & \multirow{2}{*}{Original} & \multicolumn{2}{c|}{ASan}   & \multicolumn{2}{c|}{ASan+CIMA} & \multicolumn{2}{c|}{ASan+CIMA+KASan} \\ \cline{3-8} 
                          &                           & Instrumented & Overhead     & Instrumented   & Overhead      & Instrumented     & Overhead          \\ \hline
                          \hline
Virtual memory usage      & 62.97MB                   & 549.38MB     & 8.72$\times$ & 557.5MB        & 8.85$\times$  & 699.91MB         & 11.11$\times$     \\ \hline
Real memory usage         & 8.17MB                    & 10.31MB      & 1.26$\times$ & 11.2MB         & 1.37$\times$  & 13.26MB          & 1.40$\times$      \\ \hline
Binary size               & 144KB                     & 316KB        & 2.19$\times$ & 324KB          & 2.25$\times$  & 324KB            & 2.25$\times$      \\ \hline
Shared library size       & 3196KB                    & 4288KB       & 1.34$\times$ & 4288KB         & 1.34$\times$  & 4288KB           & 1.34$\times$      \\ \hline
\end{tabular}
}
\end{table*}

\begin{table*}[htb]
\centering
\caption{Memory usage overheads of the full-stack memory-safety for Open-SecUTS}
\label{table:kasan_secuts_memory_usage}
\resizebox{\textwidth}{!}{%
\begin{tabular}{|l|l|c|c|c|c|l|l|}
\hline
\multirow{2}{*}{Category} & \multirow{2}{*}{Original} & \multicolumn{2}{c|}{ASan}   & \multicolumn{2}{c|}{ASan+CIMA} & \multicolumn{2}{l|}{ASan+CIMA+KASan} \\ \cline{3-8} 
                          &                           & Instrumented & Overhead     & Instrumented   & Overhead      & Instrumented     & Overhead          \\ \hline
                          \hline
Virtual memory usage      & 56.37MB                   & 489.29MB     & 8.68$\times$ & 490.6MB        & 8.70$\times$  & 615.33MB         & 10.92$\times$     \\ \hline
Real memory usage         & 8.76MB                    & 9.81MB       & 1.12$\times$ & 10.21MB        & 1.17$\times$  & 10.62MB          & 1.21$\times$      \\ \hline
Binary size               & 136KB                     & 288KB        & 2.12$\times$ & 296KB          & 2.18$\times$  & 296              & 2.18$\times$      \\ \hline
Shared library size       & 3196KB                    & 4288KB       & 1.34$\times$ & 4288KB         & 1.34$\times$  & 4288             & 1.34$\times$      \\ \hline
\end{tabular}
}
\end{table*}
\section{Related work}\label{related_work}



\paragraph{Cyber-Physical Systems and Memory Safety} 

Previously, we proposed to study the tolerability of a secure compilation in the context of CPS ~\cite{eyasu_cybericps}. In that work, we consider only memory error detection at user-space, does not consider mitigation resiliency against availability attacks and has a preliminary evaluation on a simulated testbed. In our other work~\cite{eyasu_essos}, we consider both kernel and user space detection, but without considering any mitigation strategy and also disregarding availability attacks. In our recent work~\cite{cima_cose}, we study dynamic instrumentation for achieving resilience against availability attacks based on memory-safety violations, but without detection of kernel-space attacks. 
In this work, we consider a full-stack memory safe compilation that is also resilient to memory-safety attacks in user-space. 


\paragraph{Generic memory safety countermeasures} Softbound~\cite{softbound} and its extension 
CETS~\cite{cets} offer a high-level memory-safety. However, 
these tools induce a very high runtime overhead (116\%), which might not be tolerable in CPS. 
Moreover, no mitigation is implemented in Softbound and CETS, hence no protection against availability attacks.  
SafeCode\cite{safecode} is also a compile-time based memory-safety tool that operates at the source code level. It instruments \emph{load} and \emph{store} instructions to prevent illegal memory accesses. 
However, SafeCode failed to prevent direct 
stack overflows toward function pointers 
and static arrays defined 
in many library functions such as fscanf(), sscanf(), sprintf() and snprintf().




\paragraph{Countermeasures based on control-flow integrity 
(CFI)}
A number of CFI-based solutions (e.g. \cite{cfi,cfi_cots,cfi_gcc,cfi_sp}) have been developed to prevent execution flow redirection attacks. 
However, these solutions have the following limitations in general: 
{\em (i)} determining the required CFG (often via a static analysis) is very difficult and requires a significant amount of memory; {\em (ii)} 
data-oriented attacks~\cite{cfi_data-attacks}, which do not divert the execution flow, cannot be detected; {\em (iii)} finally, these solutions do not implement mitigation strategies against the attacks. Consequently, the applicability of CFI-based solutions is limited in a CPS environment.

\paragraph{Memory safety and availability}  
Rinard et al. \cite{failure_oblivious} 
implemented ``failure-oblivious computing'' that allows a vulnerable program to continue its execution even in the presence of memory errors. This is accomplished via the following techniques.
Systematically fabricated values are returned for invalid memory reads, and all invalid memory writes are simply discarded. But, this approach has several limitations. Firstly, providing fabricated values to the invalid memory reads might not be always acceptable since it could result in an undesirable outcomes to the system. 
Secondly, the ``failure-oblivious computing'' 
approach is designed only against buffer-overflow vulnerabilities, hence other critical vulnerabilities, such as dangling pointers and memory leaks, are not covered. 
Finally, this approach was designed only for Servers and Desktop computers and its applicability in the context of CPS is not validated.

In summary, to the best of our knowledge, 
there is no any prior works that develops and evaluates full-stack memory-safety 
in the light of hard real-time constraints and physical-state resiliency imposed in CPS. 

\section{Conclusion}\label{conclusion}


In this research, we explored the applicability of strong countermeasures against memory-safety attacks in CPS, covering both the user-space and kernel-space attack surfaces. Moreover, we enforced a resilient mitigation strategy with a focus on availability. In particular, to evaluate efficiency of our proposed full-stack countermeasure, the induced performance overhead (both the average-case and worst-case overhead) is evaluated against the real-time constraints of the two CPS under test. 

As our experimental results revealed, the proposed full-stack countermeasure is efficient and effective enough in detecting and mitigating memory-safety attacks in a CPS environment. 
As a compile-time tool, our full-stack memory-safety enforcement is dependent on the availability of the source code. Therefore, binary instrumentation with such solutions can be considered as a future work. A resilient mitigation strategy for the kernel-space is also left as a future work.


\bibliographystyle{splncs04}
\bibliography{main.bib}


\end{document}